\documentclass[preprint,showpacs,preprintnumbers,amsmath,amssymb]{revtex4}

\usepackage{graphicx}
\usepackage{dcolumn}
\usepackage{bm}

\newcommand{\qed}{\nobreak \ifvmode \relax \else
      \ifdim\lastskip<1.5em \hskip-\lastskip
      \hskip1.5em plus0em minus0.5em \fi \nobreak
      \vrule height0.75em width0.5em depth0.25em\fi}

\begin{document}

\preprint{}

\title{A Concise Formula for Generalized Two-Qubit Hilbert-Schmidt Separability Probabilities}
\author{Paul B. Slater}%
\email{slater@kitp.ucsb.edu}
\affiliation{%
University of California, Santa Barbara, CA 93106-4030\\
}%
\date{\today}

\begin{abstract}
We report major advances in the research program initiated in "Moment-Based Evidence for Simple Rational-Valued Hilbert-Schmidt Generic $2 \times 2$ Separability Probabilities" ({\it J. Phys. A}, {\bf 45}, 095305 [2012]). A highly succinct separability probability function $P(\alpha)$ is put forth, yielding  for generic (9-dimensional) 
two-rebit systems, 
$P(\frac{1}{2}) = \frac{29}{64}$, (15-dimensional) two-qubit systems,  
$P(1) = \frac{8}{33}$ and  (27-dimensional) two-quater(nionic)bit systems,  
$P(2)=\frac{26}{323}$. This particular form of $P(\alpha)$ 
was obtained by Qing-Hu Hou by applying Zeilberger's algorithm ("creative telescoping")  to a fully equivalent--but considerably more  complicated--expression containing six 
$_{7}F_{6}$ hypergeometric functions (all with argument 
$\frac{27}{64} =(\frac{3}{4})^3$). That hypergeometric form itself had been 
obtained using systematic, high-accuracy probability-distribution-reconstruction computations. These employed 7,501 determinantal moments of partially transposed $4 \times 4$ density matrices, parameterized by  
$\alpha = \frac{1}{2}, 1, \frac{3}{2}, 2,\ldots,32$. From these computations, exact rational-valued
separability probabilities were discernible. The (integral/half-integral) sequences of 32 rational values, then,  served as input to the Mathematica FindSequenceFunction command,
from which the initially obtained  hypergeometric form of $P(\alpha)$ emerged.
\end{abstract}


\maketitle

\section{Introduction}
Our study will be devoted to addressing the fundamental quantum-information-theoretic
problem, first apparently, explicitly discussed by {\.Z}yczkowski, Horodecki, Lewenstein and Sanpera (ZHSL) \cite{ZHSL} in their highly-cited 1998 paper, "Volume of the set of separable states" \cite{ZHSL}. They gave  "three main reasons of importance"--philosophical, practical and physical--for examining such problems (cf. \cite{singh}).) Specifically, we will address the problem raised in \cite{ZHSL} of what proportion (that is, "separability probability") of quantum states are separable/disentangled \cite{RFWerner}. We endow the (generalized two-qubit) states, to which we confine our attention here, with the Hilbert-Schmidt (Euclidean/flat) metric and its accompanying measure \cite{szHS,ingemarkarol}. It is certainly also of interest to study the problem 
posed by ZHSL in alternative--but perhaps even more challenging analytically--settings, in particular that of the Bures (minimal monotone) metric/measure \cite{szBures,ingemarkarol,slaterC,slaterJGP,osipov,ye,BuresHilbert}.

We do report an apparent resolution of the ZHSL separability-probability problem in the generalized two-qubit Hilbert-Schmidt context, in terms of the titular "concise formula", which we will denote by $P(\alpha)$. Though we still lack a fully rigorous  argument for its validity, the formula strongly appears to fulfill the indicated role, while manifesting  important mathematical (random matrix theory \cite{dumitriu,tomsovic},\ldots) and physical (quantum entanglement \cite{tomsovic,ZHSL,ingemarkarol}) properties.
Thus, we have 
\begin{equation} \label{Hou1}
P(\alpha) =\Sigma_{i=0}^\infty f(\alpha+i),
\end{equation}
where
\begin{equation} \label{Hou2}
f(\alpha) = P(\alpha)-P(\alpha +1) = \frac{ q(\alpha) 2^{-4 \alpha -6} \Gamma{(3 \alpha +\frac{5}{2})} \Gamma{(5 \alpha +2})}{3 \Gamma{(\alpha +1)} \Gamma{(2 \alpha +3)} 
\Gamma{(5 \alpha +\frac{13}{2})}},
\end{equation}
and
\begin{equation} \label{Hou3}
q(\alpha) = 185000 \alpha ^5+779750 \alpha ^4+1289125 \alpha ^3+1042015 \alpha ^2+410694 \alpha +63000 = 
\end{equation}
\begin{displaymath}
\alpha  \bigg(5 \alpha  \Big(25 \alpha  \big(2 \alpha  (740 \alpha
   +3119)+10313\big)+208403\Big)+410694\bigg)+63000.
\end{displaymath}

A reader, equipped with any standard contemporary mathematical language
programming package (Maple, Mathematica, Matlab,\ldots), 
can readily verify that (to arbitrarily 
high-precision [hundreds/thousands of digits]),  quite remarkably (but not yet formally proven \cite{mathoverflow}), 
$P(0)=1,P(\frac{1}{2})=\frac{29}{64},P(1)=\frac{8}{33}$ and 
$P(2) =\frac{26}{323}$ (Figs.~\ref{fig:HypergeometricFormula1} and {fig:HouGraph}). In terms of the physical implications of the formula, we  find compelling evidence that $P(\alpha)$  yields the separability probability \cite{ZHSL}--with respect to Hilbert-Schmidt measure--of generalized two-qubit states, where, in particular $\alpha =0, \frac{1}{2}, 1, 2$ correspond to classical, rebit, qubit and quater(nionic)bit states, respectively. 

We will indicate below the multistep procedure by which the particular concise form of $P(\alpha)$ presented above was obtained. 
This process depended upon, first,  the 
derivation \cite{MomentBased} of (hypergeometric-based) formulas for the moments of  probability distributions over the determinants of partially transposed density matrices, 
followed by the 
estimation (using a certain Legendre-polynomial-based probability-distribution-reconstruction procedure \cite{Provost}) from those moments of cumulative (over the separability interval) probabilities.  Then, $\alpha$-parameterized sequences of these cumulative probabilities were analyzed to extract the underlying structure captured by 
$P(\alpha)$. This initially took a relatively complicated hypergeometric form (Fig.~\ref{fig:HypergeometricFormula1}), from which the concise formula above was subsequently derived (Figs.~\ref{fig:HouProg} and \ref{fig:HouProg2}) by Qing-Hu Hou using Zeilberger's algorithm 
\cite{doron}.

\subsection{Background}

The underpinning, predecessor paper \cite{MomentBased}--addressing the relatively 
long-standing $2 \times 2$ separability probability question \cite{ZHSL,
slaterJGP,slaterqip,slaterA,slaterC,slaterPRA,slaterPRA2,slaterJGP2,pbsCanosa,slater833} (cf. \cite{sz1,sz2,ye})--consisted largely of two sets of analyses. The first set  was concerned with establishing formulas for the bivariate determinantal product moments 
$\left\langle \left\vert \rho^{PT}\right\vert ^{n}\left\vert \rho\right\vert
^{k}\right\rangle ,k,n=0,1,2,3,\ldots,$ with respect to Hilbert-Schmidt (Euclidean/flat) measure \cite[sec. 14.3]{ingemarkarol} \cite{szHS}, of generic (9-dimensional) two-rebit and (15-dimensional) two-qubit density matrices ($\rho$). Here 
$\rho^{PT}$ denotes the partial transpose of the $4 \times 4$ density matrix $\rho$. Nonnegativity of the determinant 
$|\rho^{PT}|$ is both a necessary and sufficient condition for separability in this $2 \times 2$ setting \cite{augusiak}.

In the second set of primary analyses in \cite{MomentBased}, the {\it univariate} determinantal moments $\left\langle \left\vert \rho^{PT}\right\vert ^{n} \right\rangle$ and $\left\langle \left ( \vert \rho^{PT}\right\vert \left\vert \rho\right\vert)^n
\right\rangle$, induced using the bivariate formulas, served as input to a Legendre-polynomial-based probability distribution reconstruction algorithm of Provost \cite[sec. 2]{Provost} (cf. \cite{vericat}). This yielded estimates of the desired separability probabilities. (The reconstructed probability distributions based on $|\rho^{PT}|$ are defined over the interval $|\rho^{PT}| \in [-\frac{1}{16},\frac{1}{256}]$, while the associated separability probabilities are the cumulative probabilities of these distributions over the nonnegative 
subinterval $|\rho^{PT}| \in [0,\frac{1}{256}]$. We note that for the fully mixed (classical) state, $|\rho^{PT}| = \frac{1}{256}$, while for a maximally entangled state, such as  a Bell state,  $|\rho^{PT}| = -\frac{1}{16}$, thus, delimiting the range of $|\rho^{PT}|$.)

A highly-intriguing aspect of the (not yet rigorously established) determinantal moment formulas obtained (by C. Dunkl) in \cite[App.D.4]{MomentBased} was that both the two-rebit ($\alpha = \frac{1}{2}$) 
and two-qubit ($\alpha = 1$) cases could be encompassed by a 
{\it single} formula, with a Dyson-index-like parameter 
$\alpha$ \cite{MatrixModels} serving to distinguish the two cases. Additionally, the results of the formula for $\alpha=2$ and $n=1$ and 2 have recently been confirmed computationally by Dunkl using the "Moore determinant" (quasideterminant) \cite{Moore,Gelfand} of $4 \times 4$ quaternionic density matrices. (However, tentative efforts of ours to verify the $\alpha=4$ 
[conjecturally, {\it octonionic} 
\cite{LiaoWangLi}, problematical] case, have not proved successful.)

When the probability-distribution-reconstruction algorithm \cite{Provost} was applied 
in \cite{MomentBased} to the two-rebit case ($\alpha=\frac{1}{2}$), employing the first 3,310 moments of $|\rho^{PT}|$, 
a (lower-bound) estimate that was 0.999955 times as large as $\frac{29}{64} \approx 0.453120$ was obtained (cf. \cite[p. 6]{advances}). 

Analogously, in the two-qubit case ($\alpha =1$), using 2,415 moments, an estimate that was 0.999997066 times as large as $\frac{8}{33} \approx 0.242424$ was derived. This constitutes an appealingly simple rational value that had previously been conjectured in a quite different (non-moment-based) form of analysis, in which "separability functions" had been the main tool employed \cite{slater833}. (Note, however, that the two-rebit separability probability conjecture of $\frac{8}{17}$, somewhat secondarily advanced in \cite{slater833}, has now been discarded in favor 
of $\frac{29}{64}$.)
Let us note, supportively, that in an extensive
Monte Carlo analysis, Zhou, Chern, Fei and Joynt obtained an estimate for this two-qubit separability probability of $0.2424 \pm 0.0002$ \cite[eq. (B7)]{joynt}. Additionally, in the very same context, Fonseca-Romero, Rinc{\'o}n and Viviescas report a compatible statistic of 
$24\%$ \cite[sec. VIII]{Fonseca-Romero}.

Further, the determinantal moment formulas advanced in 
\cite{MomentBased} were then applied with $\alpha$ set equal to 2. This appears--as the indicated recent (Moore determinant) computations of Dunkl 
show--to correspond to the generic 27-dimensional set of quaternionic density matrices \cite{andai,adler}. Quite remarkably, a separability probability estimate, based on 2,325 moments, that was 0.999999987 times as large as $\frac{26}{323} \approx 0.0804954$ was found. 
\section{Outline of Present Study}
In the present study, we extend these three (individually-conducted) moment-based analyses in a more systematic, thorough manner, {\it jointly} embracing the sixty-four integral and half-integral values $\alpha =\frac{1}{2}, 1, \frac{3}{2}, 2,\ldots, 32$. We do this by accelerating, for our specific purposes, the Mathematica probability-distribution-reconstruction program of Provost \cite{Provost},  in a number of ways. Most significantly, we make use of the three-term recurrence relations for the Legendre polynomials. Doing so obviates the need to compute each successive higher-degree Legendre polynomial {\it ab initio}.

\begin{figure}
\includegraphics{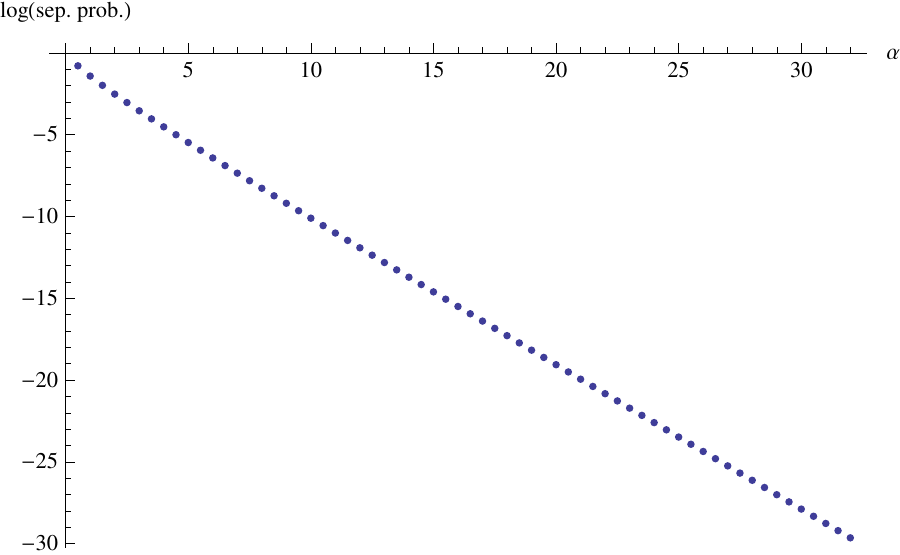}
\caption{\label{fig:ListPlotLogEstimates}Logarithms of generalized separability probability estimates, based on 7,501 Hilbert-Schmidt moments of $|\rho^{PT}|$, as a function of the Dyson-index-like parameter $\alpha$}
\end{figure}

\begin{figure}
\includegraphics{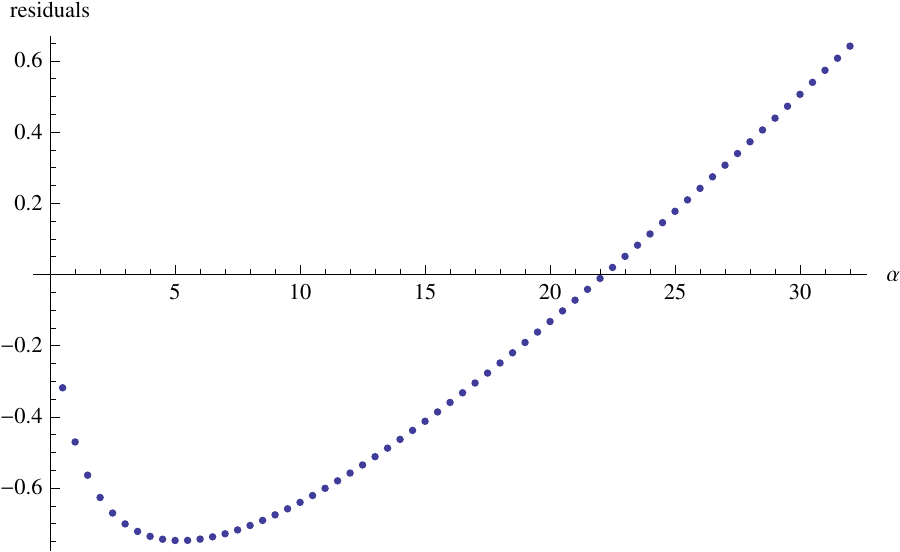}
\caption{\label{fig:EnlargedResiduals}Residuals from linear fit to logarithms of generalized separability probability estimates}
\end{figure}

In this manner, we were able to obtain--using exact computer arithmetic
throughout--"generalized" separability probability estimates based on 7,501 moments for $\alpha = \frac{1}{2}, 1, \frac{3}{2},\ldots,32$.
In Fig.~\ref{fig:ListPlotLogEstimates} we plot the logarithms of the resultant sixty-four separability probability estimates (cf. \cite[Fig. 8]{MomentBased}), which fall close to the line $-0.9464181889 \alpha$.
In Fig.~\ref{fig:EnlargedResiduals} we show the residuals from this linear fit.

\begin{figure}
\includegraphics{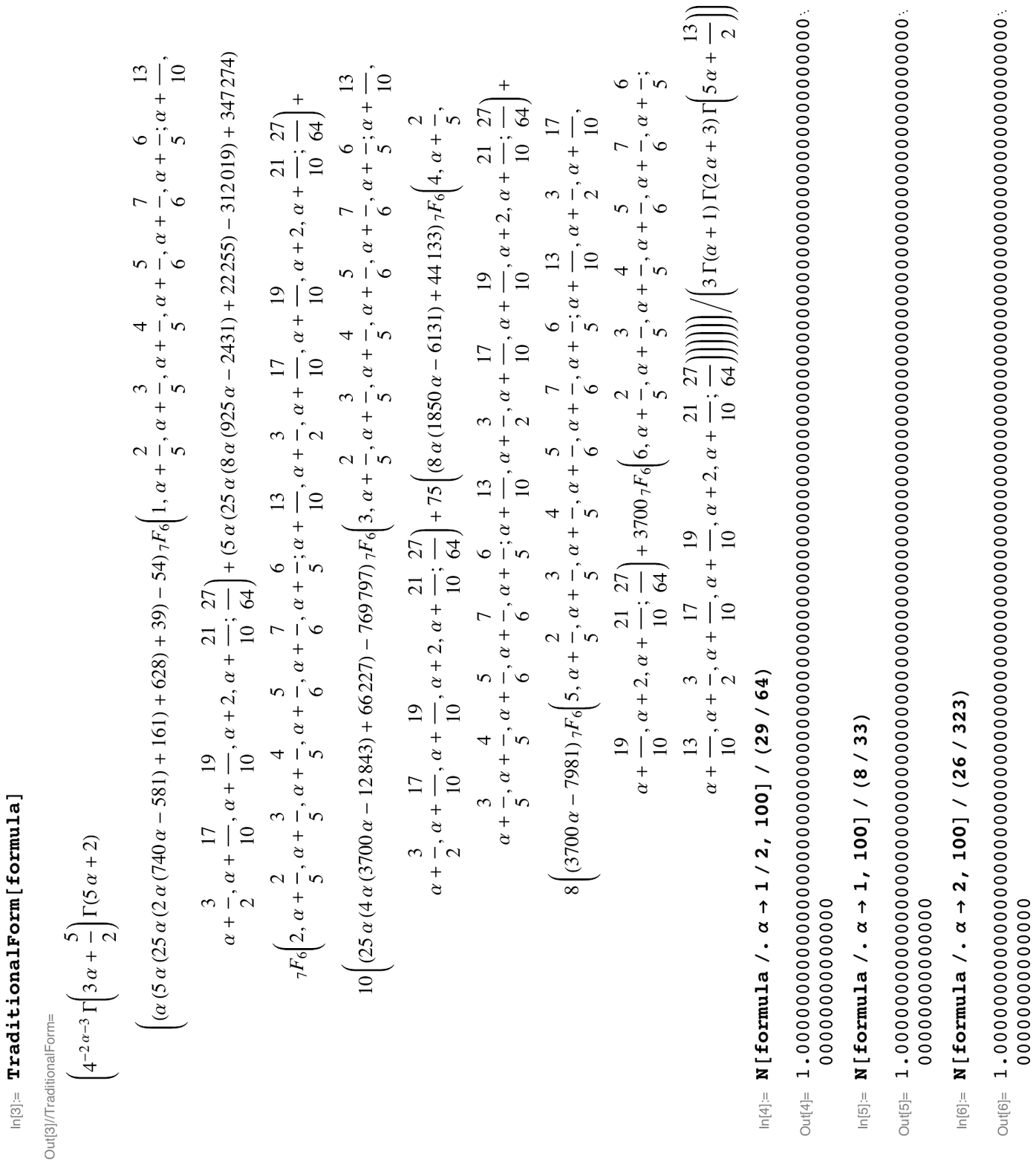}
\caption{\label{fig:HypergeometricFormula1}Hypergeometric formula $P(\alpha)$ for Hilbert-Schmidt generic $2 \times 2$ {\it generalized} separability probabilities and evidence that it reproduces the basic three (real [$\alpha = \frac{1}{2}$], complex [$\alpha = 1$] and quaternionic [$\alpha = 2$]) conjectures of 
$\frac{29}{64}, \frac{8}{33}$ and $\frac{26}{323}$}
\end{figure}

In Fig.~\ref{fig:HypergeometricFormula1} we  present a hypergeometric-function-based formula, together with striking supporting evidence
for it, that appears to succeed in uncovering the functional relation
($P(\alpha)$) 
underlying the entirely of these sixty-four generalized separability probabilities. 
Further, in 
(\ref{strikingresults}), and the immediately preceding text, 
we list a number of remarkable values yielded by this hypergeometric 
formula for values of $\alpha$ other than the basic sixty-four (half-integral and integral) values from which we have started.

Then, we are able to report--with the assistance of Qing-Hu Hou--a striking  condensation of the lengthy expression presented in Fig.~\ref{fig:HypergeometricFormula1}, that is, the titular "concise formula"
(eqs. (\ref{Hou1})-(\ref{Hou3})).

Some additional computational results of interest are presented in the Appendix.
\section{New Results}
\subsection{The three basic (rebit, qubit, quaterbit) conjectures revisited} \label{threebasics}
\subsubsection{$\alpha=\frac{1}{2}$--the two-rebit case}
In \cite{MomentBased}, a lower-bound estimate of the two-rebit separability probability was obtained, with the use of the first 3,310 moments of $|\rho^{PT}|$. It was 0.999955 times as large as
$\frac{29}{64} \approx 0.453120$. With the indicated use, now, of 7,501 moments, the figure increases to 
0.999989567. This outcome, thus, fortifies our previous conjecture.
\subsubsection{$\alpha=1$--the two-qubit case}
In \cite{MomentBased}, a lower-bound  estimate of the two-qubit separability probability was obtained, with the use of the first 2,415 moments of $|\rho^{PT}|$, that was 0.999997066 times as large as
$\frac{8}{33} \approx 0.242424$ (cf. \cite[eq. (B7)]{joynt}). Employing 7,501 moments, this figure increases to 0.99999986. 

\subsubsection{$\alpha=2$--the  quaternionic case}
In \cite{MomentBased}, a lower-bound  estimate of the (presumptive) quaternionic separability probability was obtained that was 0.999999987 times as large as $\frac{26}{323} \approx 0.0804954$, using the first 2,325 moments of $|\rho^{PT}|$. Based on 7,501 moments, this figure increases, quite remarkably still, to 0.999999999936.

\subsection{Generalized separability probability hypergeometric formula}

A principal motivation in undertaking the analyses reported here--in addition, to further scrutinizing the three specific conjectures reported in \cite{MomentBased}--was to uncover the functional relation
underlying the curve in Fig.~\ref{fig:ListPlotLogEstimates} (and/or its original non-logarithmic counterpart).

Preliminarily, let us note that the {\it zeroth}-order approximation (being independent of the particular value of $\alpha$) provided by the Provost Legendre-polynomial-based probability-distribution-reconstruction algorithm is simply the {\it uniform} distribution over the interval
$|\rho^{PT}| \in [-\frac{1}{16},\frac{1}{256}]$. The corresponding zeroth-order separability probability estimate is the cumulative probability of this distribution over the nonnegative subinterval $[0,\frac{1}{256}]$, that is, $ \frac{1}{256}/(\frac{1}{16} +\frac{1}{256}) =\frac{1}{17} \approx 0.0588235$. So, it certainly appears that speedier convergence (sec.~\ref{threebasics}) of the algorithm occurs for separability probabilities, the true values of which are initially close to 
$\frac{1}{17}$ (such as $\frac{26}{323} \approx 0.0804954$ 
in the quaternionic case).
Convergence also markedly increases as $\alpha$ increases.

It appeared, numerically, that the generalized separability probabilities for integral and half-integral values of $\alpha$ were rational values 
(not only $\frac{29}{64}, \frac{8}{33}, \frac{26}{323}$, for the three specific values $\alpha = \frac{1}{2}, 1, 2$ of original focus). With various computational tools and search strategies based upon emerging mathematical properties, we were able to advance additional,  seemingly plausible conjectures as to the exact values for $\alpha=3, 4, \ldots,32$, as well.
(We inserted many of our high-precision numerical estimates 
into the search box
on the Wolfram Alpha website--which then indicated likely candidates for corresponding rational values.) 

We fed this sequence of thirty-two conjectured rational numbers into the FindSequenceFunction command of Mathematica. (This command "attempts to find a simple function that yields the sequence $a_i$ when given successive integer arguments," but apparently can succeed with rational arguments, as well.) To our considerable satisfaction, this produced a generating formula (incorporating a diversity of hypergeometric functions of the $_{p}F_{p-1}$ type, $p=7,\ldots,11$, {\it all} with argument $z =\frac{27}{64}= (\frac{3}{4})^3$) for the sequence (cf. 
\cite[eq. (11)]{FussCatalan}). (Let us note that $z^{-\frac{1}{2}} = \sqrt{\frac{64}{27}}$ is the "residual entropy for square ice" \cite[p. 412]{finch} (cf. \cite[eqs.[(27), (28)]{ckksr}. An analogous appearance of $\frac{27}{64}$ occurs in a hypergeometric ["Ramanujan-like"] summation for  $\frac{16 \pi^2}{3}$  of J. Guillera \cite{guillera}. In a private communication, he remarked that the value $z =\frac{27}{64}$ appears to frequently occur in hypergeometric identities, and that this appears to have some modular
or modular-like origin.). In fact, the Mathematica command succeeds using only the first twenty-eight conjectured rational numbers, but no fewer--so it seems fortunate, our computations were so extensive.)

However, the formula produced by the Mathematica command was quite cumbersome in nature (extending over several pages of output). With its use, nevertheless, we were able to convincingly
generate rational values for {\it half}-integral $\alpha$ (including the two-rebit $\frac{29}{64}$ conjecture), also fitting our corresponding half-integral thirty-two numerical estimates exceedingly well.
(Let us strongly emphasize that the hypergeometric-based formula 
was initially generated using {\it only} the integral values of $\alpha$. The process was fully reversible, and we could first employ the half-integral results
to generate the formula--which then--seemingly perfectly fitted the integral values.)

At this point, for illustrative purposes, let us list the first ten half-integral and ten integral rational values (generalized separability probabilities), along with their approximate numerical values.

\begin{equation}
\begin{array}{cc}
 
\begin{array}{cccc}
 \text{$\alpha $ =} & \frac{1}{2} & \frac{29}{64} & 0.453125 \\
\end{array}
 & 
\begin{array}{cccc}
 \text{$\alpha $ =} & 1 & \frac{8}{33} & 0.242424 \\
\end{array}
 \\
 
\begin{array}{cccc}
 \text{$\alpha $ =} & \frac{3}{2} & \frac{36061}{262144} & 0.137562 \\
\end{array}
 & 
\begin{array}{cccc}
 \text{$\alpha $ =} & 2 & \frac{26}{323} & 0.0804954 \\
\end{array}
 \\
 
\begin{array}{cccc}
 \text{$\alpha $ =} & \frac{5}{2} & \frac{51548569}{1073741824} & 0.0480083 \\
\end{array}
 & 
\begin{array}{cccc}
 \text{$\alpha $ =} & 3 & \frac{2999}{103385} & 0.0290081 \\
\end{array}
 \\
 
\begin{array}{cccc}
 \text{$\alpha $ =} & \frac{7}{2} & \frac{38911229297}{2199023255552} & 0.0176948 \\
\end{array}
 & 
\begin{array}{cccc}
 \text{$\alpha $ =} & 4 & \frac{44482}{4091349} & 0.0108722 \\
\end{array}
 \\
 
\begin{array}{cccc}
 \text{$\alpha $ =} & \frac{9}{2} & \frac{60515043681347}{9007199254740992} &
   0.00671852 \\
\end{array}
 & 
\begin{array}{cccc}
 \text{$\alpha $ =} & 5 & \frac{89514}{21460999} & 0.00417101 \\
\end{array}
 \\
 
\begin{array}{cccc}
 \text{$\alpha $ =} & \frac{11}{2} & \frac{71925602948804923}{27670116110564327424} &
   0.0025994 \\
\end{array}
 & 
\begin{array}{cccc}
 \text{$\alpha $ =} & 6 & \frac{179808469}{110638410169} & 0.00162519 \\
\end{array}
 \\
 
\begin{array}{cccc}
 \text{$\alpha $ =} & \frac{13}{2} &
   \frac{3387374833367307236269}{3324546003940230230441984} & 0.0010189 \\
\end{array}
 & 
\begin{array}{cccc}
 \text{$\alpha $ =} & 7 & \frac{191151001}{298529164591} & 0.000640309 \\
\end{array}
 \\
 
\begin{array}{cccc}
 \text{$\alpha $ =} & \frac{15}{2} &
   \frac{124792688228667229196729}{309485009821345068724781056} & 0.000403227 \\
\end{array}
 & 
\begin{array}{cccc}
 \text{$\alpha $ =} & 8 & \frac{1331199762}{5232880523393} & 0.000254391 \\
\end{array}
 \\
 
\begin{array}{cccc}
 \text{$\alpha $ =} & \frac{17}{2} &
   \frac{407557367133399293946182513}{2535301200456458802993406410752} & 0.000160753 \\
\end{array}
 & 
\begin{array}{cccc}
 \text{$\alpha $ =} & 9 & \frac{74195568677}{729345064647247} & 0.000101729 \\
\end{array}
 \\
 
\begin{array}{cccc}
 \text{$\alpha $ =} & \frac{19}{2} &
   \frac{1338799759394288468677657208071}{20769187434139310514121985316880384} &
   0.0000644609 \\
\end{array}
 & 
\begin{array}{cccc}
 \text{$\alpha $ =} & 10 & \frac{730710456538}{17868447453498669} & 0.0000408939 \\
\end{array}
 \\
\end{array}
\end{equation}

To simplify the cumbersome (several-page) output yielded by the Mathematica FindSequenceFunction command, we employed certain of the "contiguous rules" for hypergeometric functions listed by 
C. Krattenthaler in his package HYP \cite{ck} (cf. \cite{bytev}). Multiple applications of the rules C14 and C18 there, together with certain gamma function simplifications suggested by C. Dunkl, led to the rather more compact formula displayed in 
Fig.~\ref{fig:HypergeometricFormula1}. 
This formula  incorporates a six-member family  ($k =1,\ldots,6$) of $_7F_6$ hypergeometric functions, differing only in the first upper index $k$,
\begin{equation} \label{family}
\, _7F_6\left(k,\alpha +\frac{2}{5},\alpha +\frac{3}{5},\alpha +\frac{4}{5},\alpha
   +\frac{5}{6},\alpha +\frac{7}{6},\alpha +\frac{6}{5};\alpha +\frac{13}{10},\alpha
   +\frac{3}{2},\alpha +\frac{17}{10},\alpha +\frac{19}{10},\alpha +2,\alpha
   +\frac{21}{10};\frac{27}{64}\right) .
\end{equation}
(The reader will note interesting sequences of upper and lower parameters (cf. \cite{zudilin}).) We are only able to, in general, evaluate the formula numerically, but then to arbitrarily high (hundreds, if not thousand-digit) precision, giving us strong confidence--despite the lack yet of a formal proof (cf. \cite{mathoverflow})--in the validity of the {\it exact} generalized separability probabilities ($\frac{29}{64}, \frac{8}{33}, \frac{26}{323}$, \ldots), that we advance.
\subsubsection{Additional interesting values yielded by the hypergeometric formula} 
Let us now apply the formula (Fig.~\ref{fig:HypergeometricFormula1}) to values of $\alpha$ other than the initial sixty-four studied. For $\alpha = 0$, the formula yields--as would be expected--the "classical separability probability" of 1. Further, proceeding in a purely formal manner (since there appears to be no corresponding  genuine probability distribution over $[-\frac{1}{16},\frac{1}{256}]$), for the {\it negative} value $\alpha = - \frac{1}{2}$, the formula  yields $\frac{2}{3}$. For $\alpha =-\frac{1}{4}$, it gives -2. Remarkably still, for $\alpha = \frac{1}{4}$, the result is clearly (to one thousand decimal places) equal to $2-\frac{34}{21 \text{agm}\left(1,\sqrt{2}\right)} =  2-\frac{17 \Gamma \left(\frac{1}{4}\right)^2}{21 \sqrt{2} \pi ^{3/2}} \approx 
0.6486993992$, where the arithmetic-geometric mean of 1 and $\sqrt{2}$ is indicated. (The reciprocal of this mean is Gauss's constant.) For $\alpha = \frac{3}{4}$, the result equals 
$2-\frac{9689 \Gamma \left(\frac{3}{4}\right)}{4420 \sqrt{\pi } \Gamma
   \left(\frac{5}{4}\right)} \approx 0.3279684732$, while for $\alpha=-\frac{3}{4}$, we have
$\frac{128}{21 \text{agm}\left(1,\sqrt{2}\right)}+2 =2+\frac{32 \sqrt{2} \Gamma \left(\frac{1}{4}\right)^2}{21 \pi ^{3/2}} \approx 7.087249321$.  For $\alpha=\frac{2}{3}$, the outcome is $2-\frac{288927 \Gamma \left(\frac{1}{3}\right)^3}{344080 \pi ^2} \approx 0.36424897456$.
Results are presented in the table 
\begin{equation} \label{strikingresults}
\left(
\begin{array}{ccc}
 \alpha  & P(\alpha ) & \text{value} \\
 -\frac{3}{4} & 2+\frac{32 \sqrt{2} \Gamma \left(\frac{1}{4}\right)^2}{21 \pi ^{3/2}} &
   7.08725 \\
 -\frac{2}{3} & 2-\frac{8 \pi }{\sqrt{3} \Gamma \left(\frac{1}{3}\right)^3} & 1.24527
   \\
 -\frac{1}{2} & \frac{2}{3} & 0.666667 \\
 -\frac{1}{3} & 2+\frac{3 \Gamma \left(\frac{1}{3}\right)^3}{4 \pi ^2} & 3.461 \\
 -\frac{1}{4} & 2 & 2 \\
 \frac{1}{4} & 2-\frac{17 \Gamma \left(\frac{1}{4}\right)^2}{21 \sqrt{2} \pi ^{3/2}} &
   0.648699 \\
 \frac{1}{3} & 2-\frac{459 \sqrt{3} \pi }{91 \Gamma \left(\frac{1}{3}\right)^3} &
   0.572443 \\
 \frac{2}{3} & 2-\frac{288927 \Gamma \left(\frac{1}{3}\right)^3}{344080 \pi ^2} &
   0.364249 \\
 \frac{3}{4} & 2-\frac{9689 \Gamma \left(\frac{3}{4}\right)}{4420 \sqrt{\pi } \Gamma
   \left(\frac{5}{4}\right)} & 0.327968 \\
\end{array}
\right) .
\end{equation}
(Let us note that the term 
$\frac{3 \Gamma \left(\frac{1}{3}\right)^3}{4 \pi ^2} \approx 1.46099848$ present in the result for 
$\alpha =-\frac{1}{3}$ is "Baxter's four-coloring constant" for a triangular lattice \cite[p. 413]{finch}.) Also, for $\alpha=-1$, we have $\frac{2}{5}$. For $\alpha=-\frac{3}{2}$, the result is $\frac{2}{3}$.
\section{Concise reformulation of $_{7}F_{6}$ hypergeometric expression 
(Fig.~\ref{fig:HypergeometricFormula1})}
\begin{figure}
\includegraphics{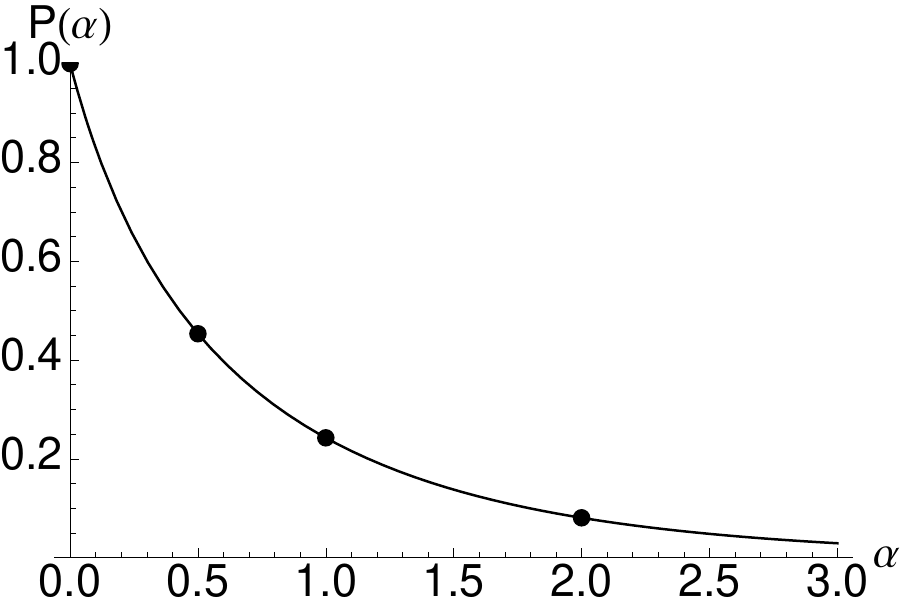}
\caption{\label{fig:HouGraph}Generalized two-qubit separability probability function $P(\alpha)$, with $P(0) =1, P(\frac{1}{2}) =\frac{29}{64}, P(1)=\frac{8}{33}, P(2)= \frac{26}{323}$ for generic classical four-level ($\alpha =0$), two-rebit ($\alpha =\frac{1}{2}$), two-qubit ($\alpha =1$) and two-quaterbit ($\alpha =2$) systems, respectively.}
\end{figure}

\begin{figure}
\includegraphics[page=1,scale=0.85]{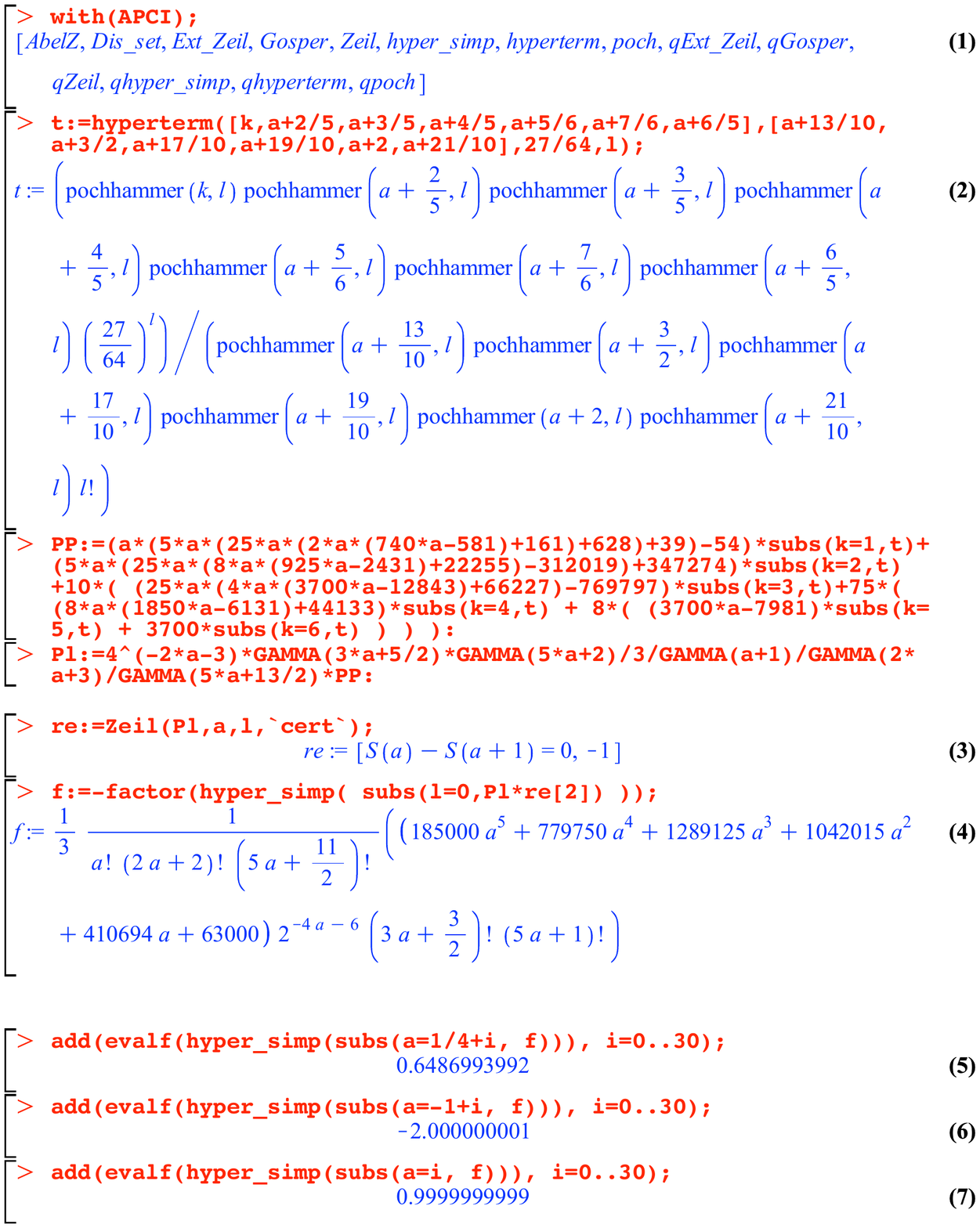}
\caption{\label{fig:HouProg}First Maple worksheet of Hou used in deriving concise form of hypergeometric formula 
(Fig.~\ref{fig:HypergeometricFormula1})}
\end{figure}

\begin{figure}
\includegraphics[page=2,scale=0.85]{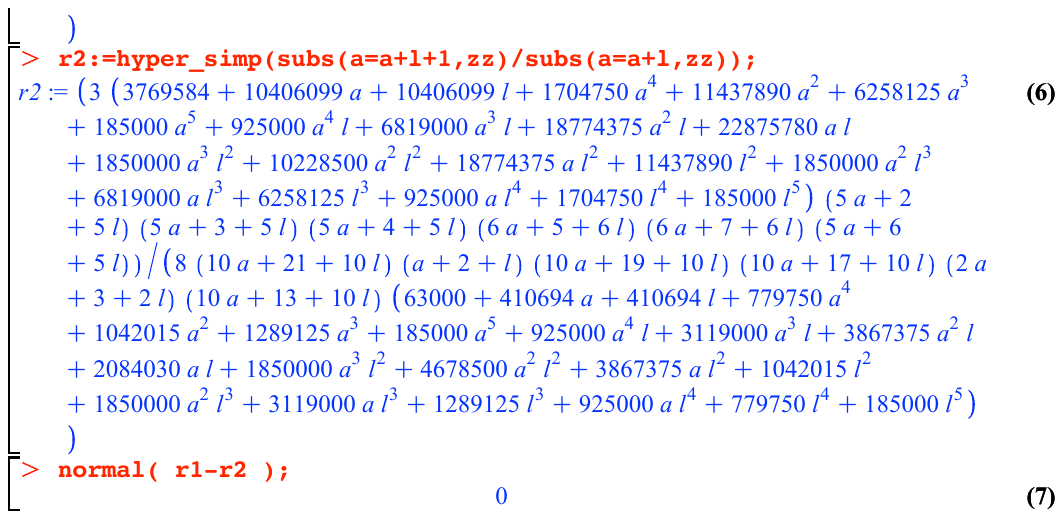}
\caption{\label{fig:HouProg2}Second Maple worksheet of Hou used in deriving concise form of hypergeometric formula 
(Fig.~\ref{fig:HypergeometricFormula1})}
\end{figure}

We had previously ourselves been unable to find an equivalent  form of $P(\alpha)$ with fewer than six hypergeometric functions (Fig.~\ref{fig:HypergeometricFormula1}). Qing-Hu Hou of the Center for Combinatorics of Nankai University, however,  was able to obtain the remarkably succinct and clearly correct results (\ref{Hou1})-(\ref{Hou3})--which he communicated to us in a few e-mail messages. (Accompanying them were two Maple worksheets indicating his calculations [Figs.~\ref{fig:HouProg} and 
\ref{fig:HouProg2}].)
Hou, first, observed that the hypergeometric-based formula for $P(\alpha)$ could be expressed as an infinite summation. Letting $P_l(\alpha)$ be the $l$-th such summand, application of Zeilberger's algorithm  \cite{doron} (a method for producing combinatorial identities, also known as "creative telescoping") yielded that 
\begin{equation} \label{referee1}
P_l(\alpha) -P_l(\alpha+1) =-P_{l+1}(\alpha) + P_l(\alpha) .
\end{equation}
(The package APCI--available at http://www.combinatorics.net.cn/homepage/hou/--was employed. In a different quantum-information context, Datta employed the algorithm to ascertain that no closed form exists for a certain series, "retarding" the evaluation of the "ratio of the negativity of random pure states to the maximal negativity for Haar-distributed states of $n$ qubits" \cite[App. A, Table I]{Datta}.)
Summing over $l$ from 0 to $\infty$, Hou found that 
\begin{equation} \label{referee2}
P(\alpha) -P(\alpha+1)=P_0(\alpha).
\end{equation}
Letting $f(\alpha) =P_0(\alpha)$, the concise summation formula (\ref{Hou1}) is obtained.
(C. Krattenthaler indicated [Krattenthaler, private communication] that these results might equally well be derived without recourse to Zeilberger's algorithm. Also, a referee expressed puzzlement at the peculiar [redundant] form of eq.~(\ref{referee1}). This appears to be an artifact arising from the particular manner in which the algorithm is applied in the proving of hypergeometric identities.)

\includegraphics[page=1,scale=0.85]{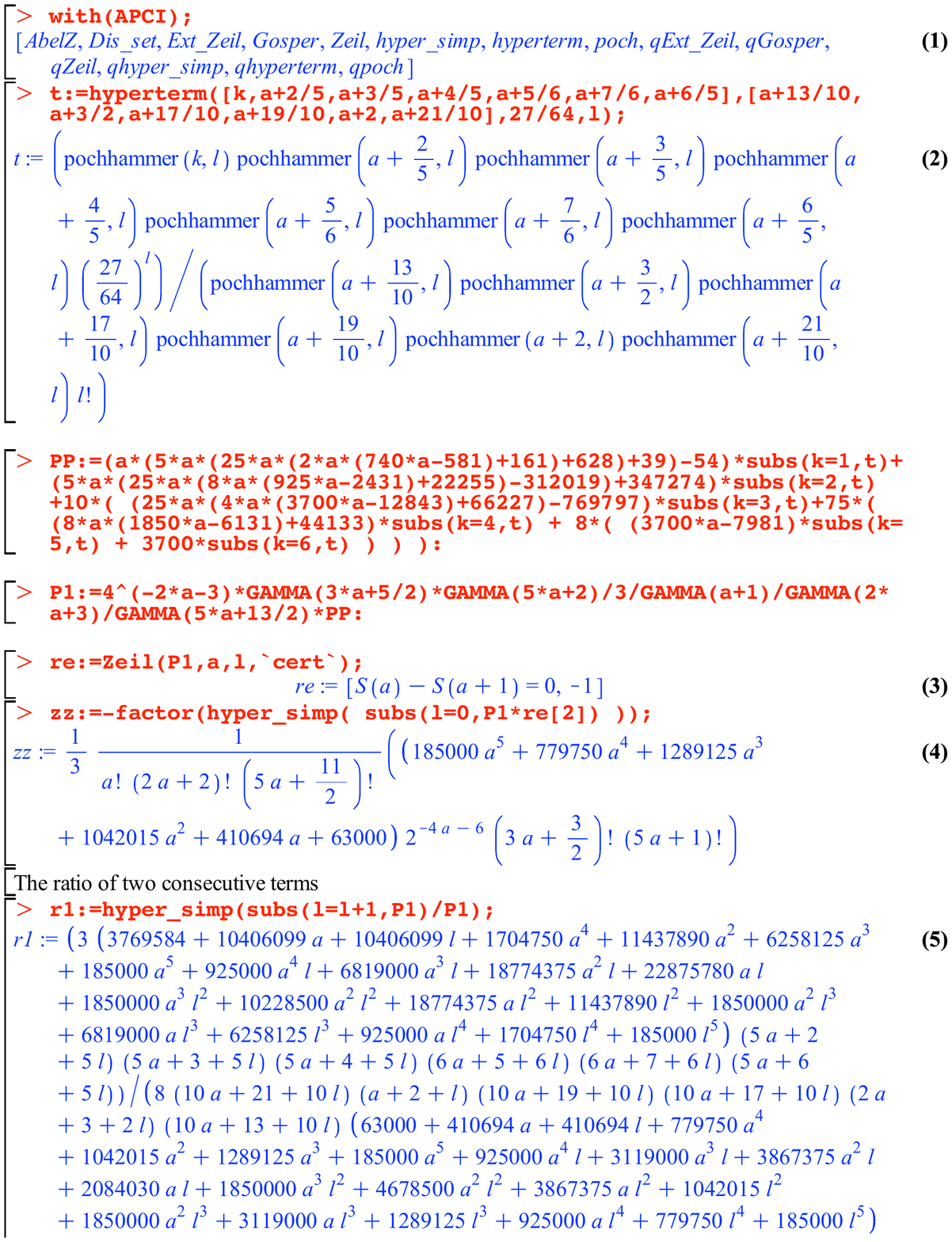}

We certainly need to indicate, however, that if we do explicitly perform the infinite summation indicated in (\ref{Hou1}), then we revert to a ("nonconcise") form of $P(\alpha)$, again containing six hypergeometric functions. Further, it appears that we can only evaluate (\ref{Hou1}) numerically--but then easily to hundreds and even thousands of digits of precision--giving us extremely high confidence in the specific rational-valued Hilbert-Schmidt separability probabilities advanced.

\section{Concluding Remarks}
There remain the important problems of formally verifying the formulas for 
$P(\alpha)$ (as well as the underlying determinantal moment formulas 
for $|\rho^{PT}|$, \ldots, in \cite{MomentBased}, employed in the probability-distribution reconstruction process), and achieving a better understanding of what these results convey regarding the geometry of quantum states \cite{ingemarkarol,avron,avron2}. Further, questions of the asymptotic behavior of the formula ($\alpha \rightarrow \infty$)  and of possible Bures metric \cite{szBures,ingemarkarol,slaterJGP,slaterqip,slaterC} counterparts to it, are under investigation \cite{BuresHilbert}.

We are presently engaged in attempting to determine further properties--in addition to the cumulative (separability) probabilities over $[0,\frac{1}{256}]$ obtained from the titular concise formula (eq. (\ref{Hou1})-(\ref{Hou3}))--of the probability distributions of $|\rho^{PT}|$ over $[-\frac{1}{16},\frac{1}{256}]$, as a function of the Dyson-index-like parameter $\alpha$.  As one such finding, it appears that the $y$-intercept (at which 
$|\rho^{PT}|=0$, that is, the separability-entanglement boundary) in the presumed quaternionic case ($\alpha=2$) is 
$\frac{7425}{34} = \frac{3^3 \times 5^2 \times 11}{2 \times 17} \approx 218.382$ \cite{slaterSuddenDeath}. (The Legendre-polynomial-based probability-distribution reconstruction algorithm  of Provost \cite{Provost} yielded an estimate 0.99999999742 times as large as $\frac{7425}{34}$, when implemented with 10,000 moments. Based also on 
10,000 moments--but with inferior convergence properties--the two-qubit [$\alpha =1$] and two-rebit [$\alpha = \frac{1}{2}$] $y$-intercepts were estimated as 389.995 (conjecturally equal to $390 = 2 \cdot 3 \cdot 5 \cdot 13$) and 502.964, respectively \cite{slaterSuddenDeath}.)

The foundational paper of {\.Z}yczkowski, Horodecki, Sanpera and Lewenstein,"Volume of the set of separable states" \cite{ZHSL} (cf. \cite{singh}), did ask for {\it volumes}, not specifically {\it probabilities}. At least, for the two-rebit, two-qubit and two-quaterbit cases, $\alpha =\frac{1}{2}, 1$ and $2$, we can readily, using the Hilbert-Schmidt volume formulas of Andai \cite[Thms. 1-3]{andai} (cf. \cite{szHS,ingemarkarol}), convert the corresponding separability probabilities to the separable volumes $\frac{29 \pi ^4}{61931520} =\frac{29 \pi^4}{2^{16} \cdot 3^3 \cdot 5 \cdot 7}$, $\frac{\pi ^6}{449513064000} = \frac{\pi^6}{2^6 \cdot 3^6 \cdot 5^3 \cdot 7^2 \cdot 11^2 \cdot 13}$ and 
$\frac{\pi ^{12}}{3914156909371803494400000} = \frac{\pi^{12}}{2^{14} \cdot 
3^{10} \cdot 5^5 \cdot 7^3 \cdot 11^2 \cdot 13 \cdot 17^2 \cdot 19^2 \cdot 23}$, respectively. The determination of separable volumes--as opposed to probabilities--for other values of $\alpha$ than these fundamental three appears to be rather problematical, however.

Let us also note the relevance of the  study of  Szarek, Bengtsson and {\.Z}yczkowski \cite{sbz}, in which they show that the convex set of separable mixed states of the 
$2 \times 2$ system is a body of  constant height. Theorem 2 of that paper, in conjunction with the results here, allows one, it would seem, to immediately deduce that  the separability probabilities of the generic  minimally-degenerate/boundary 8-, 14-, and 26-dimensional two-rebit, two-qubit, and two-quaterbit states are one-half (that is, $\frac{29}{128}, \frac{4}{33}$ and $\frac{13}{323}$) the separability probabilities of their generic non-degenerate counterparts.
\section{Appendix--Exact values of derivatives of $P(\alpha)$}
\subsection{Succeeding deriviatives at $\alpha =0$}
The first derivative of $P(\alpha)$ evaluated at (the classical case) 
$\alpha =0$ is 
-2, while the second derivative is $40 - 20 \zeta{(2)} = 40 -\frac{10 \pi^2}{3} \approx 7.10132$. (The third derivative was computed as -43.7454236566749417600.)
\subsection{First derivatives at $\alpha =1, 2 \ldots$, {\it et al}}
The first derivative of $P(\alpha)$  at $\alpha = -\frac{1}{2}$ is 
$-\frac{80}{3}$ and at $\alpha = \frac{1}{2}$ is $\frac{1}{384} (917-984 \log (2)) \approx 0.611831$, and -2 at $\alpha=0$, as previously mentioned. We have also been able to determine rational values of $P(\alpha)$ for $\alpha =1, 2, \ldots, 97$. We list the first seven of these. (The Mathematica command FindSequenceFunction, however, did not succeed in this instance in generating an underlying function for this sequence of 97 rational numbers--although, of course, one can be directly obtained from our explicit form of $P(\alpha)$.)
\begin{equation} \label{derivativeresults}
\left(
\begin{array}{ccc}
 \alpha  &  P'(\alpha) \\
 1 & 
   -\frac{130577}{457380}  \approx -0.285489\\
2 & -\frac{3177826243}{37595998440}  \approx -0.0845257\\
3 & -\frac{3598754002551529}{124409677632540300} \approx  -0.0289266 \\
4 & -\frac{943222153906869801499}{89625168823088671652880} \approx -0.0105241\\
5 & -\frac{7745868905935978063871447}{1956135029605259737354520400} \approx  -0.00395978 \\
6 & -\frac{163704960709243940550573265691777}{107569184582725029279135417408286275} \approx -0.00152186 \\
7 & -\frac{124555275071579876642057723808475761407}{209867628485254931732709294271962333917
   400} \approx -0.000593494 \\
\end{array}
\right) .
\end{equation}

\begin{acknowledgments}
I would like to express appreciation to the Kavli Institute for Theoretical
Physics (KITP)
for computational support in this research, and to Christian Krattenthaler, Charles F. Dunkl, Michael Trott and Jorge Santos for their expert advice, as well as to Qing-Hu Hou for his insights and permission to present his Maple worksheets. Further, I thank a number of referees/editors for their constructive suggestions.
\end{acknowledgments}

\bibliography{Concise100}

\end{document}